\documentclass[runningheads]{llncs}
\usepackage{graphicx}
\usepackage[T1]{fontenc}
\usepackage{verbatimbox,lipsum}
\usepackage{subcaption}
\usepackage[utf8]{inputenc}
\usepackage{listings}

\begin{document}
\title{Federating Scholarly Infrastructures with GraphQL}
%Network for Persistently Identification of Research Articles in Knowledge Graph
\author{Muhammad Haris\inst{1}\orcidID{0000-0002-5071-1658} \and
Kheir Eddine Farfar\inst{2}\orcidID{0000-0002-0366-4596} \and
Markus Stocker\inst{2}\orcidID{0000-0001-5492-3212} \and
Sören Auer\inst{2,1}\orcidID{0000-0002-0698-2864}
}

\authorrunning{Haris et al.}

\institute{L3S Research Center, Leibniz University Hannover 30167, Hannover, Germany\\
 \email {haris@l3s.de} \and
TIB---Leibniz Information Centre for Science and Technology, Germany
\email{\{kheir.farfar, markus.stocker,auer\}@tib.eu}}

\maketitle              % typeset the header of the contribution

\begin{abstract} 
A plethora of scholarly knowledge is being published on distributed scholarly infrastructures. Querying a single infrastructure is no longer sufficient for researchers to satisfy information needs. We present a GraphQL-based federated query service for executing distributed queries on numerous, heterogeneous scholarly infrastructures (currently, ORKG, DataCite and GeoNames), thus enabling the integrated retrieval of scholarly content from these infrastructures. Furthermore, we present the methods that enable cross-walks between artefact metadata and artefact content across scholarly infrastructures, specifically DOI-based persistent identification of ORKG artefacts (e.g., ORKG comparisons) and linking ORKG content to third-party semantic resources (e.g., taxonomies, thesauri, ontologies). This type of linking increases interoperability, facilitates the reuse of scholarly knowledge, and enables finding machine actionable scholarly knowledge published by ORKG in global scholarly infrastructures. In summary, we suggest applying the established linked data principles to scholarly knowledge to improve its findability, interoperability, and ultimately reusability, i.e., improve scholarly knowledge FAIR-ness.

\end{abstract}
\section{Introduction}
\label{s:introduction}
Scholarly articles are static unstructured text documents~\cite{kuhn2016decentralized} and datasets are published in a plethora of formats~\cite{Hendler_2014} with heterogeneous metadata on diverse repositories. It is thus difficult to interlink the heterogeneous collection of scholarly knowledge artefacts. Yet, complex information needs rely on retrieval from multiple infrastructures. However, querying multiple scholarly infrastructures and integrating retrieved information manually is a laborious and time-consuming task \cite{6666765,Schwarte2011FedXOT}. This problem motivates the need for unified access to and integrated retrieval from numerous, heterogeneous scholarly infrastructures. In other words, we need approaches that support formulating complex information needs via a single endpoint and retrieving integrated answers in a distributed manner \cite{Schwarte2011FedXOT}. As an example, consider the following scenario: A researcher wants to discover all research outputs (articles) published under a particular grant that have impact (high number of citations) as well as significant results (in a statistical sense of $p < .001$). It is currently impossible to formulate such an information need as a single query.

Towards this goal, we leverage the Open Research Knowledge Graph\footnote{\url{https://orkg.org}} (ORKG) ~\cite{orkg} and present a GraphQL-based endpoint\footnote{https://www.orkg.org/orkg/graphql} to access its data i.e., machine actionable descriptions of scholarly knowledge and knowledge comparisons. Furthermore, we build on this endpoint and propose a GraphQL-based federated endpoint that allows unified access to data from other scholarly infrastructures. Thus, we propose that the machine actionable scholarly knowledge published in ORKG can be leveraged along with scholarly content of other scholarly infrastructures to answer complex user queries on bibliographic metadata, article content, or both.
Specifically, we propose the following generic approaches:

\begin{enumerate}
    \item \textit{Federated access} to scholarly infrastructures to retrieve and integrate the fragmented scholarly content via a single endpoint. Here, we leverage the DataCite PID Graph\footnote{https://api.datacite.org/graphql} \cite{Fenner_Aryani_2019} and the GeoNames\footnote{\url{https://www.geonames.org/}} REST API, and link them virtually with ORKG using shared metadata.
    \item \textit{Methods for enabling federated access}
    \begin{enumerate}
        \item \textit{Persistently identifying and publishing} artefacts by leveraging DOI services. This enables cross-walking between artefact metadata and artefact content and the discovery of content. Here, we leverage DataCite\footnote{\url{https://datacite.org}} DOI-based identification of ORKG comparisons.
        \item Linking content to existing, third-party semantic resources (e.g., taxonomies, thesauri, ontologies) to improve the interoperability and reusability of scholarly knowledge. Here, we leverage GeoNames to enrich the description of locations in ORKG.
    \end{enumerate}
\end{enumerate}

We thus address the following research question: How can existing, heterogeneous scholarly infrastructures be federated to support complex (meta)data-driven analysis?

%\begin{itemize}
%	\item How can federated scholarly infrastructures be leveraged to support different stakeholder groups in satisfying their diverse information needs.
%\item How can federated scholarly infrastructures be leveraged to find semantically similar content with a single user query.
%    \textbf{RQ1}: How can existing, heterogeneous scholarly infrastructures be federated to support complex data-driven analysis?
    %\item How to ensure unified access to scholarly content published by heterogeneous scholarly infrastructures to support complex data-driven analysis?
    %\item How can structured scholarly knowledge be linked to (existing) semantic resources with unambiguous and formal meaning?
	%\item How can machine actionable scholarly knowledge be made findable in global scholarly infrastructures?
%\end{itemize}

\section{Related Work}
\label{s:related-work}

\paragraph{Persistent identifiers} Persistent identifiers (PID) are used to uniquely and persistently identify research articles, software, datasets and other digital artefacts, as well as people and physical objects such as samples~\cite{paskin2010digital}. They were introduced to decouple the identifier from the location on the Web of a digital resource and is thus an approach to address the volatility of locations, which is an issue for the stable reference (e.g., citation) required in the scholarly record, including in print material. 

Several organizations provide services to persistently identify research artefacts. Most prominently, \textit{Crossref} and \textit{DataCite} identify research articles and datasets, respectively, by means of DOI while ORCID~\cite{haak} enables the persistent identification of researchers. There exist a number of emerging identification schemes including for organizations by the \textit{Research Organization Registry} (ROR), samples by the \textit{International Geo Sample Number}\footnote{https://www.igsn.org} (IGSN), and instruments~\cite{Stocker_instruments}. Persistent identifiers thus enable the unambiguous and stable reference of research artefacts and contextual entities. A structured ORKG comparison of these persistent identifier systems for scholarly content can be found in Auer et al.~\cite{Soren}.

Since persistent identifiers have associated metadata, information about artefacts exists independently of the identified artefact. This metadata layer is typically standardized and supports the findability and accessibility of artefacts as well as enables opportunities for metadata linking and sharing among scholarly infrastructures~\cite{meadowsbuildingblocks}. The importance of persistent identification of artefacts used in research is stressed by the FAIR data principles~\cite{wilkinson2016fair}. Persistent identification is applied and adapted to numerous entity types beyond articles and datasets and their implementation is considered essential for research infrastructures. Richards et al.~\cite{richards2011beginner} discussed the persistent identification of datasets; Stocker et al.~\cite{Stocker_instruments} of instruments; Farjana et al.~\cite{FARJANA2016161} of geometric and topological entities. Bellini et al.~\cite{bellini2012interoperability} presented an interoperability framework for PID systems. Their approach is to ontologically refine the metadata sets contained in the PID records, which allows to set general information for different PID systems.

\paragraph{Semantic resources for scholarly knowledge} Ontologies are a solution for formal description of content with domain knowledge and for combining data from multiple sources~\cite{ding2007using} in data integration and classification~\cite{Salatino2018Classifying,ontologyTC}. For the scholarly domain, several ontologies have been proposed for numerous disciplines, including computer science~\cite{csontology2018,cyberSecurityOntology,ontologySE}, immunology~\cite{asiaee2015framework}, maritime research~\cite{ontology_maritime}, construction management~\cite{ontologyCM} and agronomy~\cite{agronomic_ontology}. Moreover, to describe biomedical content semantically, different bio-ontologies such as the gene ontology~\cite{Geneontology}, protein ontology~\cite{Proteinontology}, among others, have been proposed. A suite of ontologies known as Semantic Publishing and Referencing (SPAR) \cite{spar_Peroni,PERONI201233} have been proposed for the creation of comprehensive metadata for all aspects of semantic publishing and referencing (e.g., document description and bibliographic references).

\paragraph{Scholarly communication} Several frameworks have been developed to improve scholarly communication. Martin et al.~\cite{Martin2020} suggested that semantic linking is an important aspect to achieve interoperability among research infrastructures (RIs). They also discussed various ways to enhance the interoperability of RIs, including semantic contextualization, enrichment, mapping and bridging. Hajra et al.~\cite{hajra2017linking} presented a way to enhance scholarly communication by linking data from different repositories. They considered bibliographic Linked Open Data (LOD) repositories to compute the semantic similarity between two resources. With the Scholix framework~\cite{scholix}, the Research Data Alliance (RDA) Publishing Data Services Working Group (PDS-WG)~\cite{data_literature2017} developed an approach for data-literature interlinking. This framework enables interoperability of metadata about the links between articles and datasets created and exchanged among publishers and data repositories as well as scholarly infrastructures, specifically DataCite, OpenAIRE and Crossref. Another approach to interlink OpenAIRE research metadata and datasets and making metadata accessible for end users was proposed by Ameri et al.~\cite{Exploiting2017}. Assante et al.~\cite{scienceAssante} introduced Science 2.0 repositories to enhance the scholarly communication workflow by overcoming the gap between publishing research articles and research lifecycle. The structured comparison of these scholarly infrastructures can be found in Haris~\cite{Haris}.

\paragraph{Federated scholarly infrastructures.} Schwarte et al. ~\cite{Schwarte2011FedXOT} proposed a framework, named FedX, which enables efficient processing of SPARQL queries on heterogeneous data sources and also demonstrated the practicability and efficiency of the proposed framework on a set of real-world queries. Similarly, Mosharraf and Taghiyareh~\cite{federatedengine} proposed a SPARQL-based federated search engine to retrieve Open Educational Resources (OERs) published on the web of data. Arya et al. ~\cite{arya} proposed a personalized federated search framework to retrieve information such as user profiles, jobs, or professional groups from diverse sources.

Several frameworks are proposed to access scholarly data in a federated manner, but they do not consider the content of artefacts while processing user queries. To the best of our knowledge, this is the first attempt towards powering federated queries by incorporating machine-readable form of scholarly artefacts so that the results can be filtered not only at the metadata level but also at the content level. %To enable such federation, we present two crucial means of linking scholarly knowledge, namely (i) DOI-based persistent identification of ORKG artefacts and (ii) linking terms to third-party semantic resources. Such linking of machine-readable data ensures its interoperability and broad findability thus, enables the federated access to ORKG content.

\section{Approach}
\label{s:methodology}
In this section, we present the approach for federating scholarly infrastructures. We cover three key aspects. First, federated access to ORKG, DataCite and GeoNames infrastructures to retrieve and integrate the fragmented scholarly content to enable complex data-driven analysis. Second, linking ORKG content with third-party semantic resources to ensure ORKG content is interoperable and reusable. Third, DOI-based persistent identification of ORKG artefacts by using DataCite services, specifically state-of-the-art ORKG comparisons, to ensure artefact findability in global scholarly infrastructures as well as the linking of these ORKG artefacts with other artefacts, specifically articles.

\subsection{Federated Access to Scholarly Infrastructures}
This section describes the proposed approach to federate scholarly infrastructures, in particular ORKG, DataCite and GeoNames. The approach leverages GraphQL as the common interface.
DataCite provides a GraphQL endpoint for the PID Graph\footnote{https://api.datacite.org/graphql}, which connects persistently identified resources from DataCite, ORCID, ROR, etc., and serves standardized metadata for these resources. Similarly, we implemented a GraphQL endpoint\footnote{https://www.orkg.org/orkg/graphql} to enable access to ORKG content. Additionally, we also integrated the GeoNames REST API\footnote{https://www.geonames.org/export/ws-overview.html} to enable fetching information regarding continent, countries and cities using the same interface. The GeoNames API allows access to geographical data such as fetching all countries of a particular continent by specifying its continent code. For example, retrieving the list of the countries which belong to Asian (AS) continent. Thus, our proposed federation enables crosswalks between metadata about artefacts (e.g., articles and datasets) with their context (e.g., people and organizations) and the content of articles. We thus propose a GraphQL-based federation that virtually integrates ORKG, the DataCite PID Graph, and GeoNames to retrieve integrated information from these scholarly infrastructures through a single search query. Arguably, these data sources can be easily extended with additional sources.

\begin{figure*}[!tb]
\centering
\includegraphics[width=\textwidth]{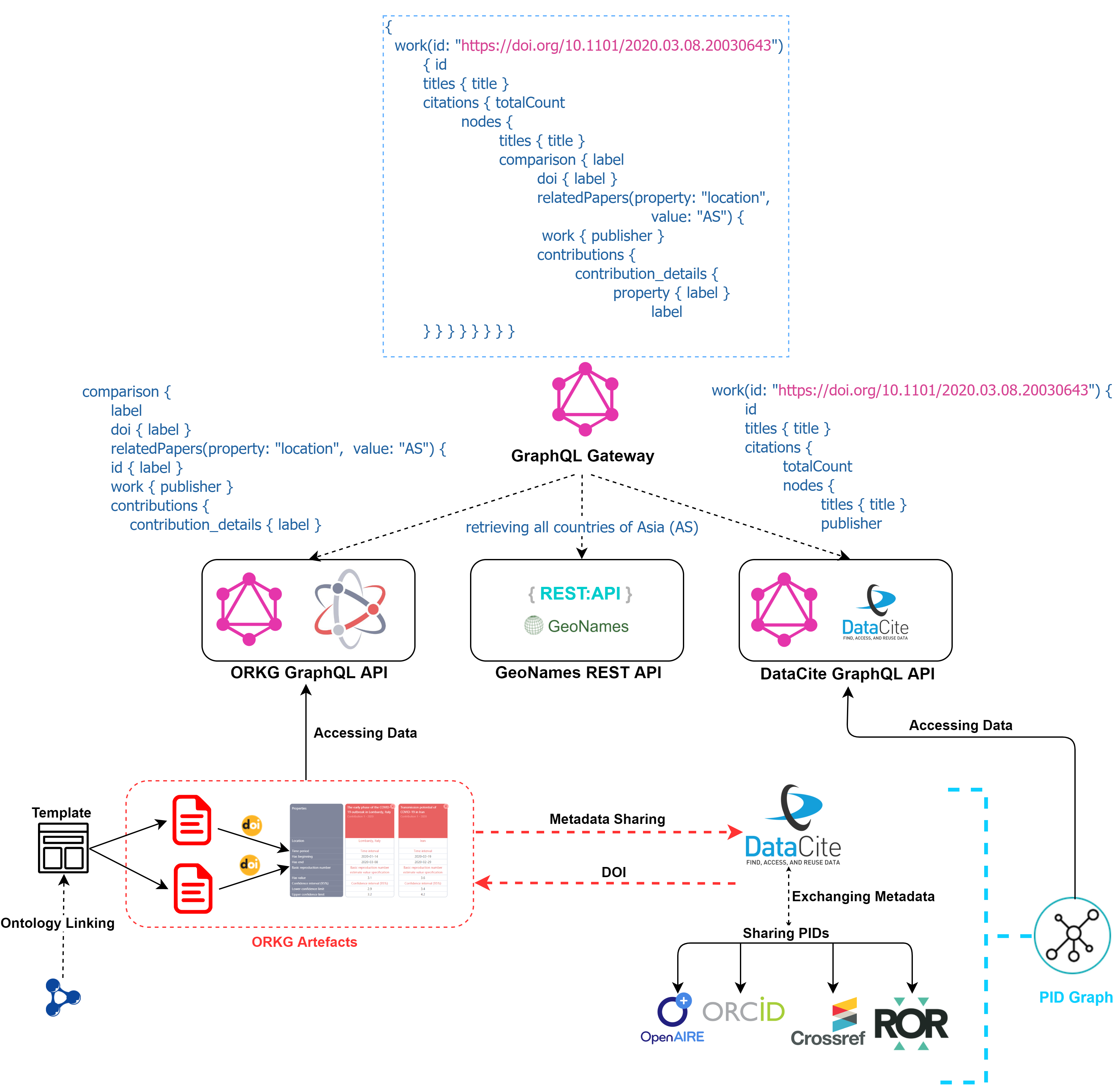}
\caption{Overview of the virtually integrated APIs of multiple scholarly infrastructures (ORKG, DataCite, and GeoNames). The figure also shows the execution of a federated query designed to retrieve from DataCite the number of citations of the paper with DOI name \texttt{10.1101/2020.03.08.20030643}, from ORKG the data of an ORKG comparison that cites the given paper, and filtering studies conducted on populations in Asia (AS). The example query thus leverages GeoNames and the proposed methods of DOI-based persistent identification of ORKG artefacts and linking content with ontologies to enable cross-walking between metadata and data served by distributed, heterogeneous scholarly infrastructures.}
\label{fig8}
\end{figure*}

Figure~\ref{fig8} illustrates the proposed federated architecture. The gateway layer virtually integrates the distinct graphs of ORKG, DataCite, and GeoNames to create a unified GraphQL endpoint, enabling the execution of a single query across these infrastructures. As shown in the figure, a single query is posed on the federated gateway in a declarative manner while query parts seamlessly execute on the respective infrastructures. The figure also depicts the underlying methods powering federated access between scholarly infrastructures, i.e., DOI-based persistent identification of machine actionable artefacts and linking the content with third-party semantic resources.
%We leverage a federation of scholarly infrastructures consisting of ORKG and DataCite to access their individual (meta)data holdings in an aggregated manner with a single query. The primary benefit of such a federation is the novel possibility of cross-walking metadata about artefacts and contextual entities (people, organizations, etc.) and scholarly knowledge published in papers, as a specific kind of artefact.

%Fig~\ref{fig1} illustrates the working of the proposed federated scholarly infrastructures. DataCite provides a GraphQL-based endpoint\footnote{https://api.datacite.org/graphql} to access its (meta)data. As shown in the figure, DataCite also shares metadata with ORCID\footnote{https://orcid.org/}~\cite{haak}, Crossref\footnote{https://www.crossref.org/} and other infrastructures. Hence, by leveraging the DataCite services, information from these infrastructures is also retrievable using the DataCite GraphQL endpoint. Similarly, ORKG also provides a GraphQL endpoint to access the machine actionable scholarly knowledge it publishes. To enable unified access to these infrastructures, their schemas are interlinked virtually on a federated gateway that supports executing distributed queries. Finally, the retrieved information is aggregated at the gateway level. 

%In the remainder of this section, we present possible usage scenarios of the proposed federated scholarly infrastructure by diverse stakeholders, namely researchers, peer reviewers, funders, and industry.

Our approach is based on a virtual integration that exhibits federation as if all scholarly infrastructures are integrated on a single endpoint. In fact, the distributed (meta)data still resides on individual infrastructures, and only the parts relevant to the query are retrieved. 
%(Meta)data located at DataCite can be fetched along with ORKG (meta)data with a single request using the common persistent identifiers. 
Hence, the proposed approach meets the goal to make global scholarly infrastructures interoperable thus serving more complex information needs. As shown in Figure \ref{fig8}, through our federated architecture, it is now possible to formulate requests with constraints on metadata about entities as well as on data, i.e. published scholarly knowledge.

A federated graph as proposed here plays an important role in enabling complex scholarly (meta)data-driven analysis. For instance, in addition to metadata analysis (e.g., citation networks) research information systems can leverage article-content (data) to power entirely new kinds of data analysis, e.g., citation networks that only include work with highly significant results, $p < .001$.

\paragraph{User Scenario.}
%TODO
A researcher reads the COVID-19 article with DOI name \texttt{10.1101/2020.03.08.20030643} and discovers that information about the virus' basic reproductive number (R0) published in the article was used in an ORKG comparison. The researcher is interested in calculating the average R0 of studies in a particular region in order to conduct regional analysis. The proposed federated endpoint enables answering such a complex query by executing the relevant parts of the query on the respective endpoints, thus enabling the (meta)data-based analysis required for the research at hand. Figure~\ref{fig8} includes the federated query that implements this user query. First, the paper is retrieved by DOI (\texttt{10.1101/2020.03.08.20030643}) on the DataCite PID graph. As this paper is cited in an ORKG comparison, the PID Graph provides us the link between the paper DOI and the DOI of the comparison. Second, using the DOI of the comparison, the query retrieves the machine actionable comparison data from ORKG. Third, by leveraging GeoNames we filter for studies in Asia. 

\begin{figure}[!tb]
\centering
\includegraphics[width=\textwidth]{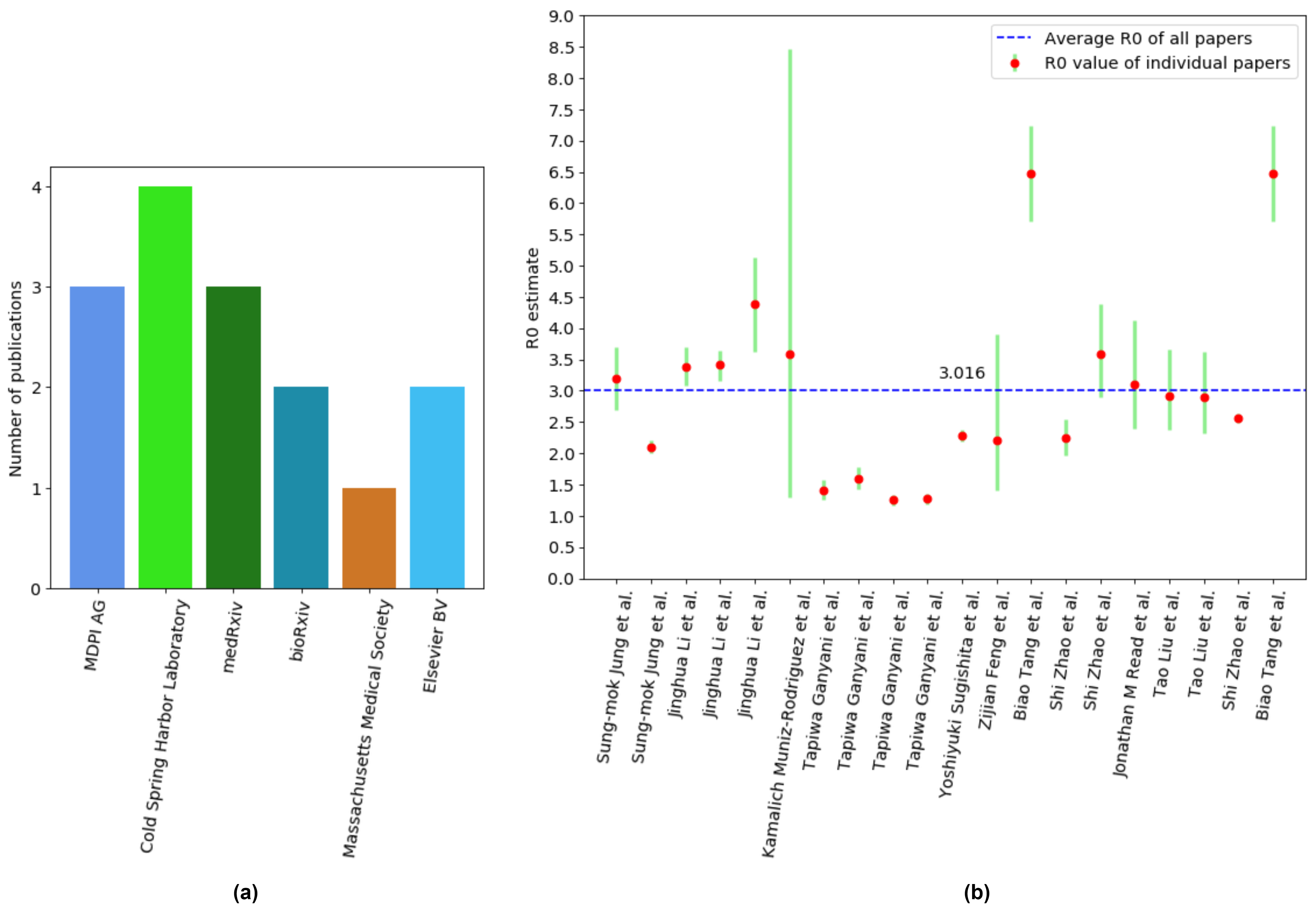}
\caption{Results of (meta)data-analysis with data retrieved using our complex user query. (a)  Metadata-based analysis: Distribution of studies across publishers. (b) Data-driven analysis: Plot of R0 estimates reported in the literature compared in ORKG and their average.}
\label{fig11}
\end{figure}

With the results obtained from such a complex query, we can perform (bibliographic) metadata-based analysis. For instance, we can compute the distribution of studies across all publishers (Figure~\ref{fig11}(a)). Such bibliographic metadata processing and analysis is a well-known and often performed activity.

The proposed approach also enables article-content (data) analysis. In our user scenario, we can for instance plot the R0 estimates reported in the literature compared in ORKG and compute the average (Figure~\ref{fig11}(b)). The computed average is 3.02. For interested readers, the (meta)data analysis discussed here is available online as a Jupyter notebook\footnote{\url{https://gitlab.com/TIBHannover/orkg/orkg-notebooks/-/blob/master/COVID-19\_R0\_meta-data\_analysis.ipynb}}.

\subsection{Methods for Linking Scholarly Knowledge}
We present two methods for linking scholarly knowledge, underlying the proposed query federation among scholarly infrastructures.
\subsubsection{Linking Semantic Resources.}
Structuring and semantically representing scholarly knowledge is generally non-trivial. The difficulty strongly depends on the granularity of the description as well as the existence of relevant and reusable schemes, among other reasons. 

To facilitate structuring scholarly knowledge, we have proposed ORKG templates\footnote{https://www.orkg.org/orkg/templates}. Their purpose is to ease creating comparable content in ORKG. For recurrent information types, e.g., time intervals or quantity values, their unit and confidence interval, templates specify the required properties and apply validation rules on value types to ensure quality and comparable data. This is similar to how SHACL~\cite{knublauch2017shapes} or ShEx~\cite{prud2014shape} set constraints and validate RDF data. Once having been specified, templates can be used to create ORKG content, in particular to describe research contributions. Templates not only ease content creation but, even more importantly, also standardize the description of scholarly knowledge in the ORKG.

Of most relevance here, templates not only support structuring scholarly knowledge but also allow for linking the terms used (classes, properties and individuals) with third-party semantic resources, thereby ensuring that ORKG content created using templates is interoperable and reusable. Hence, to make structured content in ORKG semantic, i.e., machine actionable, we developed a mechanism that supports the efficient linking of ORKG terms with third-party semantic resources. These resources can be accessed and linked using two generic data exchange approaches: REST API and SPARQL. ORKG provides run time support to find and access semantic resources while adding and curating content in ORKG.

While specifying a template it is thus possible to link the template to a particular class, either user-defined or defined by an existing semantic resource. Whenever a particular template is used for describing research contributions in ORKG, the semantics specified by the template are applied to the created content. We present the advantages of linking semantic resources with the following examples.

\begin{figure}[tb]
\centering
\includegraphics[width=\columnwidth]{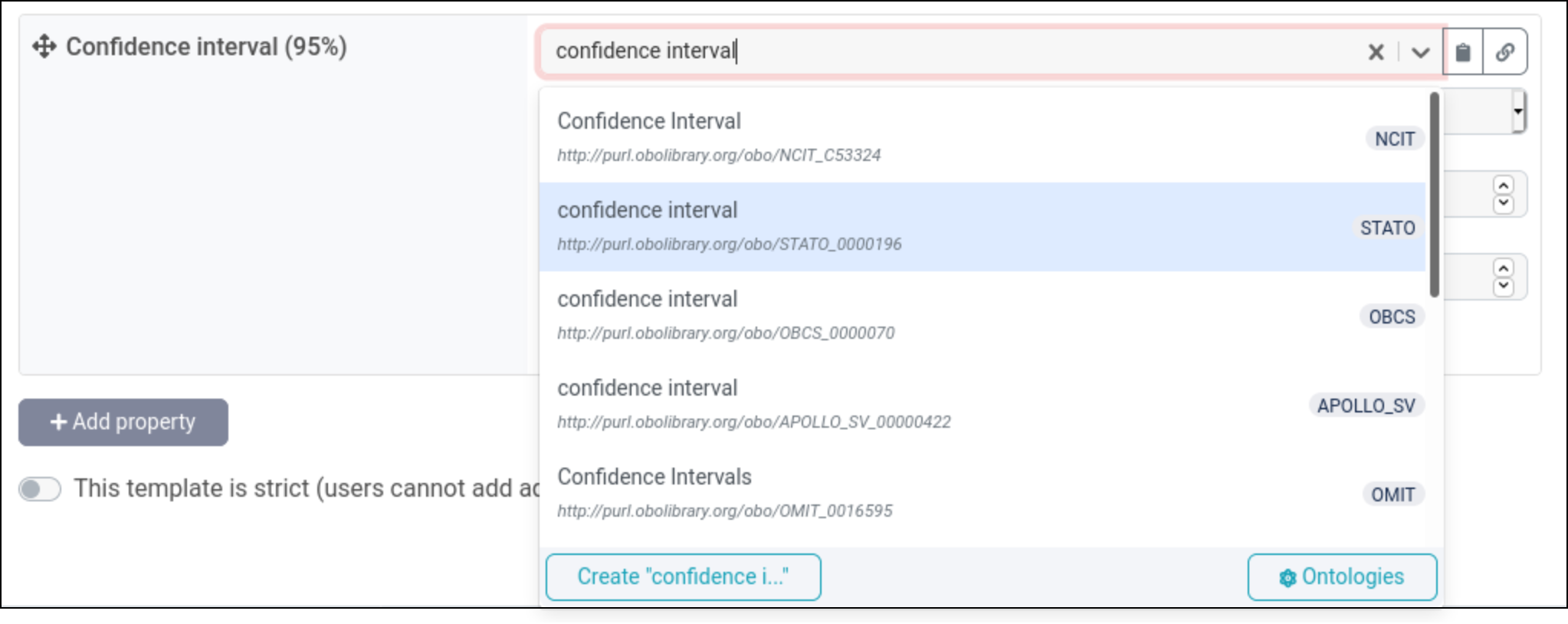}
\caption{Lookup and linking externally defined classes using the EBI-OLS semantic resource.} 
\label{fig2}
\end{figure}

\begin{figure}[t!]
\centering
\includegraphics[width=\columnwidth]{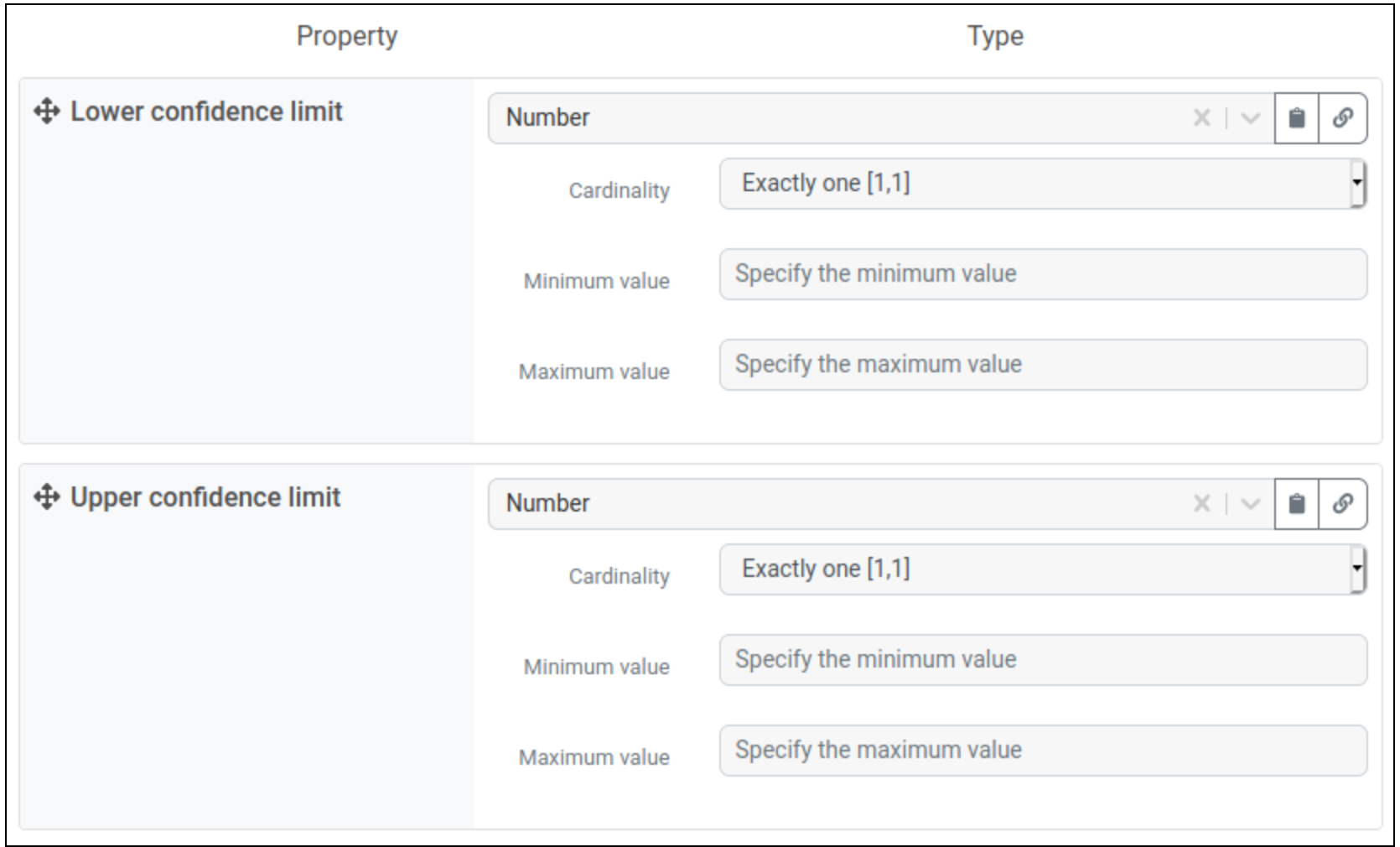}
\caption{Confidence interval template specification with constraints on property values.} \label{fig3}
\end{figure}

\textbf{Example 1.} In this example, we leverage the EMBL-EBI Ontology Lookup Service~\cite{ci2010ontology} (OLS) and its REST API. While specifying templates, it is possible to lookup classes of semantic resources served by the OLS and use these to specify value ranges.  For instance, if different studies mention the confidence interval of measurements in experiments, a respective template can refer to the Statistical Methods Ontology (STATO). Whenever a paper contribution description uses that template to specify properties and values, these will be automatically associated with the ontology. The creation of a confidence interval class and definition of constraints for template properties is shown in Figure~\ref{fig2} and Figure~\ref{fig3}.

\begin{figure}[t!]
\centering
\includegraphics[width=7cm]{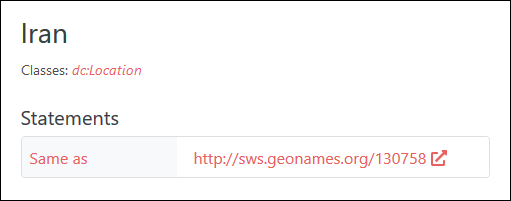}
\caption{ORKG location with automatically created \texttt{Same as} relation to the corresponding GeoNames resource.} \label{fig4}
\end{figure}

\textbf{Example 2.} In this example, we leverage the GeoNames gazetteer~\cite{vatant2012GeoNames} and its REST API to enable the lookup of GeoNames resources when specifying locations in ORKG content. We define a template using the Dublin Core (DC) Location ontology to represent locations. When using this template in ORKG research contribution descriptions, the class \texttt{DCLocation}\footnote{\url{https://www.orkg.org/orkg/class/DCLocation}} is automatically associated with locations entered as values. While typing location names, a list is fetched from the GeoNames Rest API and the user can select the country from the list, thus creating a dynamic location resource in ORKG which refers to the corresponding GeoNames resource using the \texttt{Same as} property (e.g., Figure~\ref{fig4} for the country of Iran).

\subsubsection{Persistent Identification and Linking of ORKG Comparisons.}
 
\begin{figure}[!b]
\centering
\includegraphics[width=\columnwidth]{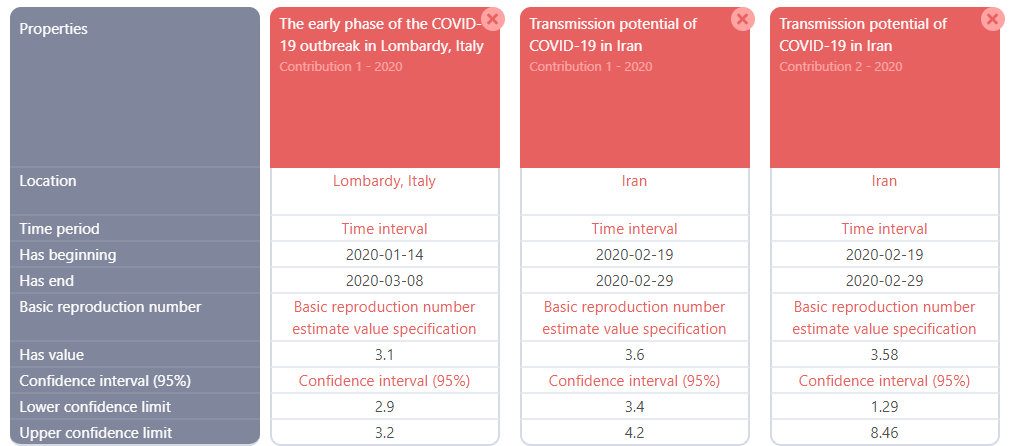}
\caption{ORKG comparison of studies estimating the COVID-19 basic reproductive number.} \label{fig6}
\end{figure}

\begin{figure}[t!]
\centering
\includegraphics[width=6cm]{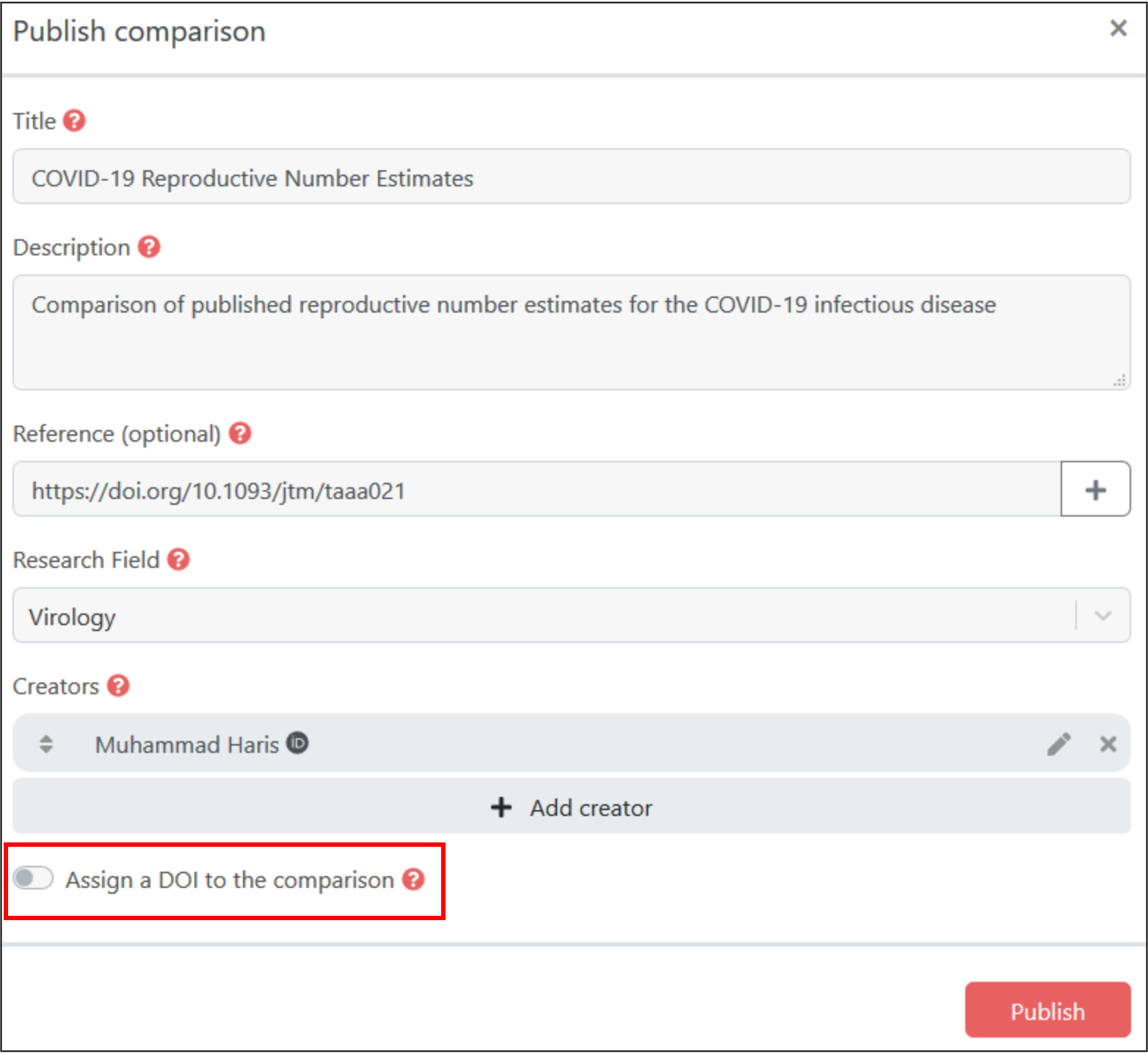}
\caption{ORKG form to persistently publish an ORKG comparison with metadata and DOI.} \label{fig5}
\end{figure}

ORKG comparisons are machine actionable tabular overviews of essential information published in the literature w.r.t. a specific research problem. They provide a condensed overview of the state-of-the-art for the respective research problem. Comparisons can be persistently stored with added metadata, including title, description, research field and creators. Figure~\ref{fig6} shows the comparison of research contributions that estimate the COVID-19 basic reproductive number whereas Figure~\ref{fig5} shows the form displayed to users to publish a comparison (including assignment of a DOI). A detailed description of ORKG comparisons can be found in Oelen et al.~\cite{survey_literature}. 

ORKG supports the DOI-based persistent identification of its comparisons to make these artefacts citeable and findable in scholarly infrastructures. A DOI is assigned to a comparison by leveraging DataCite services and publishing metadata following the DataCite metadata schema\footnote{\url{https://schema.datacite.org}} through its REST API\footnote{\url{https://support.datacite.org/docs}}.

ORKG ensures that the published metadata associated with the comparison DOI includes links between the DOI of the comparison and DOIs of the compared literature. Similarly, other PIDs (e.g., contributor ORCID IDs and organization IDs) are also included in the metadata. This metadata is shared with DataCite, which itself shares relevant elements with other infrastructures. By leveraging this existing mechanism, we can ensure that ORKG content and thus machine actionable scholarly knowledge is findable in global scholarly infrastructures. This sharing of metadata enables the cross-walking between ORKG and other scholarly infrastructures based on shared persistent identifiers. 
Listing~\ref{listing1} shows the metadata shared when registering a comparison DOI with DataCite and illustrates the linking between the DOI of the comparison and the DOIs of compared literature. The most important metadata elements are:
%\vspace{-1em}
\begin{itemize}
	\item \textit{identifier}: Represents the DOI of the comparison.
	\item \textit{creators}: Has two sub elements: \textit{creatorName} and \textit{nameIdentifier}, which represent the name and the ORCID of the creator respectively. If the ORCID is included, DataCite exchanges metadata about the link between the comparison DOI and the creator ORCID with ORCID, thus acknowledging the creator's contribution to ORKG in the contributor's ORCID record.
	\item \textit{subject}: The research field the comparison belongs to, e.g., database systems or information systems.
	\item \textit{resourceType}: Since comparisons are tabular overviews of scholarly knowledge, we consider comparisons to be datasets.
	\item \textit{relatedIdentifiers}: Links a comparison with other related resources, in particular articles. Articles included in a comparison are linked here by their DOI (if available). The sub element \textit{relatedIdentifierType} specifies the type of the related identifier (e.g., DOI, URL, etc.) whereas \textit{relationType} specifies the kind of relation this resource has with the related resource. In our case, compared literature is referenced by the comparison.
\end{itemize}

\begin{lstlisting}[language=XML, caption=Metadata used to publish an ORKG comparison with DataCite., basicstyle=\ttfamily\small, numbers=left, xleftmargin=0.5cm, numbersep=4pt, showspaces=false, showstringspaces=false, label=listing1]
<?xml version="1.0" encoding="UTF-8"?>
<resource xmlns="http://datacite.org/schema/kernel-4"
  xmlns:xsi="http://www.w3.org/2001/XMLSchema-instance"
  xsi:schemaLocation="http://datacite.org/schema/kernel-4" 
  "http://schema.datacite.org/meta/kernel-4.3/metadata.xsd">

<identifier identifierType="DOI">10.48366/r44930</identifier>

<creators>
  <creator>
    <creatorName nameType="Personal">Haris, Muhammad
    </creatorName>
    <nameIdentifier schemeURI="http://orcid.org/"
                    nameIdentifierScheme="ORCID"> 
      0000-0002-5071-1658</nameIdentifier>
  </creator>
</creators>

<titles>
  <title xml:lang="en">COVID-19 Reproductive Number Estimates
  </title> 
</titles>
<publisher xml:lang="en">Open Research Knowledge Graph
</publisher>
<publicationYear>2020</publicationYear>

<subjects>
  <subject xml:lang="en">Virology</subject>
</subjects>

<language>en</language>
<resourceType resourceTypeGeneral="Dataset">Comparison
</resourceType>

<relatedIdentifiers>  
  <relatedIdentifier relationType="References"
        relatedIdentifierType="DOI">
        10.1101/2020.03.08.20030643</relatedIdentifier>
</relatedIdentifiers>
<descriptions>
  <description descriptionType="Abstract">
  Comparison of published reproductive number estimates 
  for the COVID-19 infectious disease</description>
</descriptions>
</resource>
\end{lstlisting}

\section{Discussion}
\label{s:discussion}
We have presented a GraphQL-based federated data retrieval system architecture that supports cross-walks between (bibliographic) metadata and article's content (data) across multiple scholarly infrastructures. We suggested that such seamless cross-walking enables formulation of complex user queries. To enable cross-walks, we also presented two crucial means of linking scholarly knowledge, namely (i) DOI-based persistent identification of ORKG artefacts and (ii) linking terms to third-party semantic resources. Our work suggests that Persistent Identifiers (PIDs) are not only interesting to persistently identify scholarly documents but can also be leveraged to persistently identify machine actionable representations of essential information contained in documents. We demonstrated this by leveraging DataCite services to persistently identify ORKG comparisons, thus making ORKG content broadly findable in global scholarly infrastructures. Moreover, software agents can also discover ORKG content using the DataCite PID Graph, which further supports fetching machine-actionable data in a federated manner based on the shared metadata and performing complex (meta)data-driven analysis.

We also demonstrated a generic approach to efficiently link ORKG terms to third-party semantic resources while specifying ORKG templates used to create linked ORKG content. Such kind of linking makes the ORKG content unambiguous, interoperable and reusable, and supports both humans and machines in knowledge discovery.

To address our research question, we virtually integrate the DataCite PID Graph and GeoNames REST API with ORKG, thus ensuring the retrieval of required information in a federated manner. In addition to classical bibliographic metadata analysis, we emphasize that using artefact bibliographic metadata in combination with artefact machine actionable content greatly empowers the federation to enable complex (meta)data-driven analysis needed in research.   
The integrated access to multiple scholarly infrastructures realizes two main advantages. First, it provides up-to-date results and, second, it provides interesting insights into the connections in research (meta)data. While state-of-the-art scholarly infrastructures provide linking among artefacts at the metadata level, the linked access to artefact content proposed here enables novel (meta)data-driven analysis currently not supported by state-of-the-art scholarly infrastructures. 

While the proposed architecture can be extended, the presented implementation is limited to a federation of  three infrastructures, namely DataCite, GeoNames and ORKG. The implementation can be extended to additional infrastructures, but doing so relies on additional software development. A further challenge is writing GraphQL queries, which can be difficult for untrained users. % The proposed federation is limited to manually querying the scholarly infrastructures and is not able to automatically provide dynamic links of relevant external data with other ORKG artefacts.

To address these limitations, we aim to enhance the scope of federated queries by virtually integrating other (scholarly) infrastructures (OpenAIRE\footnote{\url{https://www.openaire.eu/}}, Wikidata\footnote{\url{https://www.wikidata.org/wiki/Wikidata:Main\_Page}}, Zenodo\footnote{\url{https://zenodo.org/}}, etc.). We also plan to develop a user interface to allow users to pose their queries in the form of facets, thus automatically generating GraphQL queries to retrieve the required results in a federated manner. Moreover, we aim to develop a dashboard that demonstrates the linking of related content (dataset, articles) residing on global scholarly infrastructures with ORKG machine-actionable artefacts (e.g., comparisons) and other entities (e.g., organizations).
%The proposed system addresses the coordination of multiple infrastructures using the metadata managed in these infrastructures. This virtual integration provides global model semantics as if all infrastructures are integrated locally. Therefore, the federated access to scholarly infrastructures provides their distinct information in an aggregated manner, thus enabling complex data-driven analysis.
%ORKG's machine-readable artefacts have proven to be crucial for searching and analyzing scholarly content to a deep extent. Our proposed federated endpoint leverages the ORKG machine-readable content to meet the diversified information needs of different stakeholders. The use of machine-readable content allows different stakeholders to find statistically significant papers along with the retrieval of other related scholarly artefacts using DataCite services. 

\section{Conclusion}
\label{s:conclusion}
Federated search is an established practice in retrieving data from disparate sources. However, to the best of our knowledge, to date no federated architecture allows scholarly metadata-to-data cross-walks including structured scholarly knowledge in the form presented here for complex (meta)data-driven analysis of scholarly knowledge. 

As the main contribution of this work, we presented a federated architecture that supports cross-walks from metadata to data of scholarly artefacts. Our implementation leverages DataCite, ORKG and GeoNames infrastructures. %The proposed federated architecture leverages the machine-actionable scholarly knowledge published in ORKG to answer complex user queries. %Our federated endpoint benefits researchers to pose complex queries to meet their diverse information needs. 

We also presented two important approaches for linking scholarly knowledge to empower the federation among scholarly infrastructures, i.e., DOI-based persistent identification of ORKG artefacts and linking content with third-party semantic resources. We demonstrated this kind of linking with an implementation in ORKG. %Therefore, leveraging the machine-readable content in federated architecture allows retrieving statistically significant information to address complex user queries.
%Linking terms to third-party semantic resources is a resource intense activity. Hence, we have proposed an implementation that enables such linking in templates thus ensuring that content created with templates is automatically linked.

Persistent identification and linking of machine actionable scholarly knowledge with related artefacts, agents, organizations, etc. is also an essential component towards linked scholarly knowledge. As global scholarly infrastructures already extensively share persistent identifier metadata, we have exploited such kind of linking to improve ORKG content findability.

As the amount of scholarly knowledge published by global scholarly infrastructures relentlessly increases, yet scholarly knowledge remains poorly machine processable, we argue that federated and retrieval with constraints on metadata and data should receive more attention in the community. We suggest that the proposed federated system can support researchers in conducting science by effectively enabling them in the formulation of complex information needs that are typical for modern science.

\section*{Acknowledgment}
This work was co-funded by the European Research Council for the project ScienceGRAPH (Grant agreement ID: 819536) and TIB--Leibniz Information Centre for Science and Technology. The authors thank Mohamad Yaser Jaradeh for his valuable input and comments.

\bibliographystyle{splncs04}
\bibliography{paper}

\begin{thebibliography}{10}
\providecommand{\url}[1]{\texttt{#1}}
\providecommand{\urlprefix}{URL }
\providecommand{\doi}[1]{https://doi.org/#1}

\bibitem{Exploiting2017}
Ameri, S., Vahdati, S., Lange, C.: Exploiting interlinked research metadata.
  pp. 3--14 (09 2017). \doi{{10.1007/978-3-319-67008-9\_1}}

\bibitem{arya}
Arya, D., Ha-Thuc, V., Sinha, S.: Personalized federated search at linkedin.
  In: Proceedings of the 24th ACM International on Conference on Information
  and Knowledge Management. p. 1699–1702. CIKM '15, Association for Computing
  Machinery, New York, NY, USA (2015). \doi{10.1145/2806416.2806615},
  \url{https://doi.org/10.1145/2806416.2806615}

\bibitem{Geneontology}
Ashburner, M., Ball, C., Blake, J., Botstein, D., Butler, H., Cherry, J.,
  Davis, A.P., Dolinski, K., Dwight, S., Eppig, J., Harris, M., Hill, D.,
  Issel-Tarver, L., Kasarskis, A., Lewis, S., Matese, J., Richardson, J.,
  Ringwald, M., Rubin, G., Sherlock, G.: Gene ontology: tool for the
  unification of biology. the gene ontology consortium. Nat Genet  \textbf{25},
   25--29 (05 2000)

\bibitem{asiaee2015framework}
Asiaee, A.H., Minning, T., Doshi, P., Tarleton, R.L.: A framework for
  ontology-based question answering with application to parasite immunology.
  Journal of biomedical semantics  \textbf{6}(1), ~31 (2015)

\bibitem{scienceAssante}
Assante, M., Candela, L., Castelli, D., Manghi, P., Pagano, P.: Science 2.0
  repositories: Time for a change in scholarly communication. D-Lib Magazine
  \textbf{21} (01 2015). \doi{10.1045/january2015-assante}

\bibitem{Soren}
Auer, S., Stocker, M.: Comparison of scholarly identifier systems (2021).
  \doi{10.48366/R73210}, \url{https://www.orkg.org/orkg/comparison/R73210}

\bibitem{bellini2012interoperability}
Bellini, E., Luddi, C., Cirinn{\`a}, C., Lunghi, M., Felicetti, A., Bazzanella,
  B., Bouquet, P.: Interoperability knowledge base for persistent identifiers
  interoperability framework. In: 2012 Eighth International Conference on
  Signal Image Technology and Internet Based Systems. pp. 868--875. IEEE (2012)

\bibitem{data_literature2017}
Burton, A., Koers, H., Manghi, P., La~Bruzzo, S., Aryani, A., Diepenbroek, M.,
  Schindler, U.: The data-literature interlinking service: Towards a common
  infrastructure for sharing data-article links. Program  \textbf{51},  75--100
  (04 2017). \doi{10.1108/PROG-06-2016-0048}

\bibitem{scholix}
Burton, A., Koers, H., Manghi, P., Stocker, M., Fenner, M., Aryani, A.,
  La~Bruzzo, S., Diepenbroek, M., Schindler, U.: The scholix framework for
  interoperability in data-literature information exchange. D-Lib Magazine
  \textbf{23} (01 2017). \doi{10.1045/january2017-burton}

\bibitem{ci2010ontology}
C\^{o}té, R., Reisinger, F., Martens, L., Barsnes, H., Vizcaino, J.,
  Hermjakob, H.: The ontology lookup service: bigger and better. Nucleic acids
  research  \textbf{38}(suppl\_2),  W155--W160 (2010)

\bibitem{ding2007using}
Ding, L., Kolari, P., Ding, Z., Avancha, S.: Using ontologies in the semantic
  web: A survey. In: Ontologies, pp. 79--113. Springer (2007)

\bibitem{FARJANA2016161}
Farjana, S.H., Han, S., Mun, D.: Implementation of persistent identification of
  topological entities based on macro-parametrics approach. Journal of
  Computational Design and Engineering  \textbf{3}(2),  161 -- 177 (2016).
  \doi{10.1016/j.jcde.2016.01.001}

\bibitem{Fenner_Aryani_2019}
Fenner, M., Aryani, A.: {Introducing the PID Graph}  (2019).
  \doi{10.5438/JWVF-8A66},
  \url{https://blog.datacite.org/introducing-the-pid-graph/}

\bibitem{haak}
Haak, L., Fenner, M., Paglione, L., Pentz, E., Ratner, H.: Orcid: A system to
  uniquely identify researchers. Learned Publishing  \textbf{25},  259--264 (10
  2012). \doi{10.1087/20120404}

\bibitem{hajra2017linking}
Hajra, A., Tochtermann, K.: Linking science: approaches for linking scientific
  publications across different lod repositories. International Journal of
  Metadata, Semantics and Ontologies  \textbf{12}(2-3),  124--141 (2017)

\bibitem{ontologySE}
Happel, H.J., Seedorf, S.: Applications of ontologies in software engineering
  (01 2006)

\bibitem{Haris}
Haris, M.: Comparison of scholarly infrastructures (2021).
  \doi{10.48366/R73195}, \url{https://www.orkg.org/orkg/comparison/R73195}

\bibitem{Hendler_2014}
Hendler, J.: Data integration for heterogenous datasets. Big data  \textbf{2},
  205--215 (12 2014). \doi{10.1089/big.2014.0068}

\bibitem{cyberSecurityOntology}
Iannacone, M., Bohn, S., Nakamura, G., Gerth, J., Huffer, K., Bridges, R.,
  Ferragut, E., Goodall, J.: Developing an ontology for cyber security
  knowledge graphs. pp.~1--4 (04 2015). \doi{10.1145/2746266.2746278}

\bibitem{orkg}
Jaradeh, M.Y., Oelen, A., Farfar, K.E., Prinz, M., D'Souza, J., Kismih\'{o}k,
  G., Stocker, M., Auer, S.: Open research knowledge graph: Next generation
  infrastructure for semantic scholarly knowledge. In: Proceedings of the 10th
  International Conference on Knowledge Capture. p. 243–246. K-CAP '19,
  Association for Computing Machinery, New York, NY, USA (2019).
  \doi{10.1145/3360901.3364435}

\bibitem{agronomic_ontology}
Jonquet, C., Dzalé-Yeumo, E., Arnaud, E., Larmande, P.: Agroportal : a
  proposition for ontology-based services in the agronomic domain  (06 2015)

\bibitem{knublauch2017shapes}
Knublauch, H., Kontokostas, D.: Shapes constraint language (shacl). W3C
  Candidate Recommendation  \textbf{11}(8) (2017)

\bibitem{kuhn2016decentralized}
Kuhn, T., Chichester, C., Krauthammer, M., Queralt-Rosinach, N., Verborgh, R.,
  Giannakopoulos, G., Ngomo, A.C.N., Viglianti, R., Dumontier, M.:
  Decentralized provenance-aware publishing with nanopublications. PeerJ
  Computer Science  \textbf{2}, ~e78 (2016)

\bibitem{Martin2020}
Martin, P., Magagna, B., Liao, X., Zhao, Z.: Semantic linking of research
  infrastructure metadata. In: Towards Interoperable Research Infrastructures
  for Environmental and Earth Sciences, pp. 226--246. Springer (2020).
  \doi{{10.1007/978-3-030-52829-4\_13}}

\bibitem{meadowsbuildingblocks}
Meadows, A., Haak, L., Brown, J.: Persistent identifiers: the building blocks
  of the research information infrastructure. Insights the UKSG journal
  \textbf{32} (03 2019). \doi{10.1629/uksg.457}

\bibitem{federatedengine}
Mosharraf, M., Taghiyareh, F.: Federated search engine for open educational
  linked data. Bull. IEEE Tech. Comm. Learn. Technol  \textbf{18}(6) (2016)

\bibitem{Proteinontology}
Natale, D., Arighi, C., Barker, W., Blake, J., Bult, C., Caudy, M., Drabkin,
  H., D'Eustachio, P., Evsikov, A., Huang, H., Nchoutmboube, J., Roberts, N.,
  Smith, B., Zhang, J., Wu, C.: The protein ontology: A structured
  representation of protein forms and complexes. Nucleic acids research
  \textbf{39},  D539--45 (10 2010). \doi{10.1093/nar/gkq907}

\bibitem{survey_literature}
Oelen, A., Jaradeh, M.Y., Stocker, M., Auer, S.: Generate fair literature
  surveys with scholarly knowledge graphs. In: Proceedings of the ACM/IEEE
  Joint Conference on Digital Libraries in 2020. p. 97–106. JCDL '20,
  Association for Computing Machinery, New York, NY, USA (2020).
  \doi{10.1145/3383583.3398520}

\bibitem{paskin2010digital}
Paskin, N.: Digital object identifier (doi) system. encyclopedia of library and
  information sciences. Tech. rep. (2010)

\bibitem{PERONI201233}
Peroni, S., Shotton, D.: Fabio and cito: Ontologies for describing
  bibliographic resources and citations. Journal of Web Semantics  \textbf{17},
   33 -- 43 (2012). \doi{https://doi.org/10.1016/j.websem.2012.08.001}

\bibitem{spar_Peroni}
Peroni, S., Shotton, D.: The spar ontologies. In: Vrande{\v{c}}i{\'{c}}, D.,
  Bontcheva, K., Su{\'a}rez-Figueroa, M.C., Presutti, V., Celino, I., Sabou,
  M., Kaffee, L.A., Simperl, E. (eds.) The Semantic Web -- ISWC 2018. pp.
  119--136. Springer International Publishing, Cham (2018)

\bibitem{prud2014shape}
Prud'hommeaux, E., Labra~Gayo, J.E., Solbrig, H.: Shape expressions: an rdf
  validation and transformation language. In: Proceedings of the 10th
  International Conference on Semantic Systems. pp. 32--40 (2014)

\bibitem{richards2011beginner}
Richards, K., White, R., Nicolson, N., Pyle, R.: A beginner’s guide to
  persistent identifiers. GBIF (2011)

\bibitem{Salatino2018Classifying}
Salatino, A., Thanapalasingam, T., Mannocci, A., Osborne, F., Motta, E.:
  Classifying research papers with the computer science ontology. In:
  International Semantic Web Conference (2018)

\bibitem{csontology2018}
Salatino, A.A., Thanapalasingam, T., Mannocci, A., Osborne, F., Motta, E.: The
  computer science ontology: A large-scale taxonomy of research areas. In:
  Vrande{\v{c}}i{\'{c}}, D., Bontcheva, K., Su{\'a}rez-Figueroa, M.C.,
  Presutti, V., Celino, I., Sabou, M., Kaffee, L.A., Simperl, E. (eds.) The
  Semantic Web -- ISWC 2018. pp. 187--205. Springer International Publishing,
  Cham (2018)

\bibitem{ontologyTC}
Sanchez-Pi, N., Martí, L., {Bicharra Garcia}, A.C.: Improving ontology-based
  text classification: An occupational health and security application. Journal
  of Applied Logic  \textbf{17},  48 -- 58 (2016).
  \doi{10.1016/j.jal.2015.09.008}, sOCO13

\bibitem{ontology_maritime}
Santipantakis, G., Kotis, K., Vouros, G.: "ontology-based data integration for
  event recognition in the maritime domain (07 2015).
  \doi{10.1145/2797115.2797133}

\bibitem{Schwarte2011FedXOT}
Schwarte, A., Haase, P., Hose, K., Schenkel, R., Schmidt, M.: Fedx:
  Optimization techniques for federated query processing on linked data. In:
  International Semantic Web Conference (2011)

\bibitem{Stocker_instruments}
Stocker, M., Darroch, L., Krahl, R., Habermann, T., Devaraju, A., Schwardmann,
  U., D'Onofrio, C., H{\"a}ggstr{\"o}m, I.: Persistent identification of
  instruments. Data Science Journal  \textbf{19},  1--12 (05 2020).
  \doi{10.5334/dsj-2020-018}

\bibitem{vatant2012GeoNames}
Vatant, B., Wick, M.: Geonames ontology. Dostupn{\'e} online:< http://www.
  geonames. org/ontology/ontology\_v3  \textbf{1} (2012)

\bibitem{wilkinson2016fair}
Wilkinson, M.D., Dumontier, M., Aalbersberg, I.J., Appleton, G., Axton, M.,
  Baak, A., Blomberg, N., Boiten, J.W., da~Silva~Santos, L.B., Bourne, P.E.,
  et~al.: The fair guiding principles for scientific data management and
  stewardship. Scientific data  \textbf{3}(1), ~1--9 (2016)

\bibitem{ontologyCM}
Zhang, S., Boukamp, F., Teizer, J.: Ontology-based semantic modeling of
  construction safety knowledge: Towards automated safety planning for job
  hazard analysis (jha). Automation in Construction  \textbf{52},  29 -- 41
  (2015). \doi{10.1016/j.autcon.2015.02.005}

\bibitem{6666765}
{Zhou}, Y., {De}, S., {Moessner}, K.: Implementation of federated query
  processing on linked data. In: 2013 IEEE 24th Annual International Symposium
  on Personal, Indoor, and Mobile Radio Communications (PIMRC). pp. 3553--3557
  (2013). \doi{10.1109/PIMRC.2013.6666765}

\end{thebibliography}
\end{document}